# Unusual band evolution and persistence of topological surface states in high-$T_C$ magnetic topological insulator

K. Hori[1], S. Souma[2,3*], C.-W. Chuang[1,4], Y. Nakata[1], K. Nakayama[1], S. Gupta[4,†], T. P. T. Nguyen[5], K. Yamauchi[5], T. Takahashi[1], F. Matsukura[4,6], F. H. Chang[7], H. J. Lin[7], C. T. Chen[7], A. Chainani[7], and T. Sato[1,2,3,8,9*]

[1]*Department of Physics, Graduate School of Science, Tohoku University, Sendai 980-8578, Japan*

[2]*Center for Science and Innovative in Spintronics, Tohoku University, Sendai 980-8577, Japan*

[3]*Advanced Institute for Materials Research (WPI-AIMR), Tohoku University, Sendai 980-8577, Japan*

[4]*Center for Innovative Integrated Electronic Systems, Tohoku University, Sendai 980-0845, Japan*

[5]*Department of Precision Engineering, Graduate School of Engineering, Osaka University, 2-1 Yamadaoka, Suita, Osaka 565-0871, Japan*

[6]*RIKEN Center for Emergent Matter Science, Wako, 351-0198*

[7]*National Synchrotron Radiation Research Center, Hshinchu, 30077, Taiwan ROC*

[8]*International Center for Synchrotron Radiation Innovation Smart (SRIS), Tohoku University, Sendai 980-8577, Japan*

[9]*Mathematical Science Center for Co-creative Society (MathCCS), Tohoku University, Sendai 980-8577, Japan*

[†]*Present address: Department of Physics, Bennett University, Greater Noida, 201310, India*

**ABSTRACT**

Understanding the mechanism of ferromagnetism in ferromagnetic topological insulators (TIs) is a key to realize exotic time-reversal-symmetry-broken quantum phases. However, electronic states relevant to the ferromagnetism are highly controversial. Here we report angle-resolved photoemission spectroscopy on $(Cr_xSb_{1-x})_2Te_3$ thin films, high-Curie-temperature ($T_C$) ferromagnetic TIs, spanning the non-doped ($T_C$=0 K) to highly-doped ($T_C$=192 K) region. We found that, upon Cr doping to $Sb_2Te_3$, the bulk valence-band valley exhibits filling-in behavior while retaining band inversion, leading to the formation of a nearly-flat band in high-$T_C$ regime and evolution from a six-petal flower to a Star-of-David Fermi surface. Despite the weakening of spin-orbit coupling with Cr doping, the Dirac-cone state persists up to the highest-$T_C$ sample, and shows a clear magnetic-gap opening below $T_C$ accompanied with an unexpected band shift, signifying its strong coupling with spontaneous ferromagnetism. The present result lays the foundation for understanding the interplay between band topology and ferromagnetism in TIs.

Coupling between topology and magnetism is a key ingredient to functionalize topological materials. Introduction of ferromagnetism (FM) into topological insulators (TIs) is essential to realize some exotic quantum phenomena such as the quantum anomalous Hall (QAH) effect[1,2] and the topological magnetoelectric effect[3], and it further leads to novel topological phases like the Weyl semimetal[4,5] and axion insulator[6]. These novel phases are characterized by the breaking of time-reversal symmetry and creation of topological edge/surface states associated with non-zero Chern number in the electronic structure under the magnetic order[1-6]. Also, the spin-momentum locking and strong spin-orbit coupling of the topological surface state (SS) make the ferromagnetic TIs promising for spintronics applications[7-12]. To explore exotic physical properties and functionalities by utilizing the coupling of topological and ferromagnetic properties[13–15], it is indispensable to realize FM with high Curie temperature ($T_C$) in TIs.

Tetradymite Bi and Sb chalcogenides are excellent candidates for this purpose and has been exclusively studied because the replacement of Bi/Sb with transition-metal elements is known to induce FM order in the crystal[16–20]. Spontaneous FM in tetradymite TIs was originally found in Fe-doped $Bi_2Te_3$ ($T_C \sim 10$ K)[16], and subsequently, the FM with higher $T_C$ (exceeding 100 K) was achieved in Cr- and V-doped $Sb_2Te_3$ (refs. 21–24).

From the observed correlation between concentration of magnetic impurities $n_{mag}$ and $T_C$, higher $n_{mag}$ is regarded to be necessary to achieve higher $T_C$ (refs. 17, 18, 21–25). On the other hand, an increase in $n_{mag}$ reduces the strength of spin-orbit coupling (SOC) and eventually triggers topological phase transition into ordinary insulator phase that is not topologically useful[25–29]. Furthermore, since the antiferromagnetism (AFM) tends to be stabilized by the short-range super-exchange coupling as in magnetic semiconductors[30,31], whether or not the ferromagnetic order can be stabilized by overcoming such antiferromagnetic interaction depends on the detailed electronic structure of material[32–34]. So far, a few mechanisms have been proposed to explain the FM in TIs[2,22,25–27,34–46], among which RKKY (Ruderman, Kittel, Kasuya, and Yosida) interaction and van Vleck mechanism [or BR (Bloembergen-Rowland) mechanism] are regarded to be prime candidates. The former is based on the exchange coupling among magnetic ions mediated by the valence-band (VB) hole carriers[22,32,35,36,41,42,44–46], whereas the latter is associated with the enhancement of magnetic susceptibility for valence electrons due to the band inversion[2,27,37,40]. However, it is still unclear how these mechanisms are responsible for the occurrence of high-$T_C$ FM[15].

Besides the stabilization of FM, it is necessary to create an energy gap at the Dirac point of the topological SS to realize the QAH effect and the topological magnetoelectric effect. However, whether or not such a Dirac gap can be visualized by spectroscopies is controversial. For example, although the Dirac gap was first reported in Fe- and Mn-doped $Bi_2Se_3$ (refs. 19, 20), its link to the FM remains elusive because the Se-based tetradymites never show spontaneous FM[47,48]. Also, although the QAH effect observed in the Te-based tetradymites is suggestive of the existence of Dirac gap[23,39,49], such a gap has not been observed to date. The existence or absence of the Dirac gap is also controversial in intrinsic magnetic TIs like MnBi$_2$Te$_4$ (refs. 50–53) and the family compounds[54–57]. All these necessitate further spectroscopic studies on the magnetic TIs showing robust bulk ferromagnetic order and high $T_C$. Here we access these essential issues by utilizing comprehensive angle-resolved photoemission spectroscopy (ARPES) on epitaxial ferromagnetic TI films (Cr$_{1-x}$Sb$_x$)$_2$Te$_3$ covering a wide range of $x$ ($x$ = 0.0–0.35) including very high-$T_C$ ($T_C$ = 192 K) region.

At first we present characterization of our (Cr$_x$Sb$_{1-x}$)$_2$Te$_3$ films. To elucidate the systematic evolution of electronic states as a function of $T_C$ by ARPES, we prepared samples with four different Cr concentrations and $T_C$'s; namely, ($x$, $T_C$) = (0.0, 0 K), (0.045, 30 K), (0.15, 98 K), and (0.35, 192 K). Representative magnetization curves as a function of temperature for $T_C$=98 K and 192 K samples as well as the plot of $T_C$ against $x$ are shown in Fig. 1a and its inset, respectively. We have confirmed that the cleaved surface contains reasonably wide terraces as shown by a typical atomic-force-microscopy (AFM) image in Fig. 1b. We have also confirmed by XMCD measurements that the surface of (Cr$_x$Sb$_{1-x}$)$_2$Te$_3$ samples exhibits FM consistent with the bulk magnetization measurements. As shown in Fig. 1c, the X-ray absorption spectrum on the Cr $L_3$ and $L_2$ edges for the $T_C$ =98 K sample measured at well below $T_C$ ($T$ = 25 K) under the magnetic field of 1T (top panel) shows an obvious difference between the positive and negative fields. Similar but relatively enhanced difference was also observed for the $T_C$=192 K sample (middle panel). This is more clearly visualized by a comparison of the XMCD intensity at $T$ = 25 K (bottom panel) which signifies overall lager magnitude for the $T_C$=192 K sample compared to the $T_C$=98 K one, consistent with the difference in the bulk magnetization values at low temperature shown in Fig. 1a.

Having established the occurrence of ferromagnetic order in the high Cr content films, we present the ARPES determination of band structure for $x = 0.0$–$0.35$. Figure 2a shows the $hv$ dependence of ARPES intensity for pristine $Sb_2Te_3$ ($x = 0.0$) measured along the $k_x$ cut ($\overline{\Gamma M}$ cut) that covers a wide **k** area surrounding the $\Gamma$ and Z point of bulk Brillouin zone (BZ) shown in Fig. 2b, c. One can recognize in the ARPES intensity at $hv = 66$ eV a dispersive band crossing the Fermi energy ($E_F$) showing linear dispersion within binding-energy ($E_B$) range of $E_F$–0.3 eV. This band is also seen at $hv = 70$ and 74 eV and is attributed to the lower branch of topological Dirac-cone SS whose Dirac point is located at above $E_F$ due to the hole-doped nature of the $Sb_2Te_3$ film[58] (note that the Dirac cone is not clearly seen at $hv = 78$–86 eV probably due to the matrix-element effects of photoelectron intensity). There also exists trivial SS at $E_B = 0.4$–0.8 eV whose energy position is robust against the $hv$ variation[59]. This SS is surrounded by the bulk VB whose bulk origin can be inferred from the obvious $hv$ dependence; the VB topped at $k_x \sim 0.35$ Å$^{-1}$ gradually moves upward on approaching $hv = 78$ eV from 66 eV, and then moves downward on further increasing $hv$. It is thus suggested that $hv = 78$ eV is well suited to trace the VB top hosting a three-dimensional (3D) hole pocket around the Z point as shown in Fig. 2c. The ARPES intensity at $hv = 78$ eV in Fig. 2a signifies a M-shaped VB with a deep valley around the $\overline{\Gamma}$ point bottomed at $E_B \sim 0.8$ eV. A sudden intensity reduction across $E_F$ due to the cut-off from the Fermi-Dirac (FD) distribution function is observed around the VB top at $k_x \sim 0.35$ Å$^{-1}$, suggesting that the VB crosses $E_F$ in line with the metallic transport property. It is noted that, despite the $E_F$ crossing, the peak position in the energy distribution curves (EDCs) always stays below $E_F$ likely due to the $k_z$ broadening effect of 3D bulk band, as reported in other TIs and p-type degenerate semiconductors[60,61].

Next we present the evolution of electronic states with Cr doping. Figure 3a shows the ARPES intensity at $T = 30$ K measured along the $k_x$ cut around the $\bar{\Gamma}$ point for the (Cr$_x$Sb$_{1-x}$)$_2$Te$_3$ films with four different $T_C$'s (0, 30, 98, and 192 K). Corresponding ARPES intensity at $E_F$ as a function of two-dimensional wave vector ($k_x$ and $k_y$) is also shown in Fig. 3b. The photon energy is fixed to $h\nu = 78$ eV to visualize systematically the evolution of VB top and bulk FS topology (note that similar **k** points in the 3D BZ can be traced by fixing $h\nu$ even for the Cr-doped sample because the $c$-axis lattice constant is nearly independent of $x$). The Fermi surface (FS) mapping for Sb$_2$Te$_3$ shown in the left panel of Fig. 3b signifies six small elongated pockets (petals) surrounding the $\bar{\Gamma}$ point associated with the $E_F$-crossing of VB. We have estimated the Fermi wave vectors (**k**$_F$'s) for the petals by tracing the **k** point at which the Fermi-edge intensity suddenly drops (Fig. 3c). We have assumed 3D ellipsoidal shape of the petals by referring to the band-structure calculations, and estimated the hole carrier concentration $n_h$ to be $\sim 2.2 \times 10^{19}$ cm$^{-3}$.

A side by side comparison of ARPES intensities in Fig. 3a indicates that spectral features become broader with increasing $x$, due to the stronger electron scattering associated with the Cr doping. The Cr doping also causes changes in the topmost VB dispersion, i.e., the valley at the $\bar{\Gamma}$ point is gradually filled in with Cr doping, and eventually the dispersion becomes nearly flat around the $\bar{\Gamma}$ point for the $T_C$=192 K sample. Such non-rigid-band evolution (for details, see Fig. S1 of Supplementary Information) can be also recognized from the FS mapping in Fig. 3b and corresponding schematic 3D FS in Fig. 3d where the six-petal FSs expand toward the $\bar{\Gamma}$ point and merge into a single large Star-of-David FS. Such an intriguing FS expansion which cannot be explained in terms of the simple spectral broadening indicates that the number of hole carriers increases with Cr doping, despite the isovalent (3+) nature of Sb and Cr ions[22,24,36].

Thus, the Cr doping likely changes the condition of crystal defects. By applying the same procedure to estimate $n_h$ from the FS volume as in $x = 0.0$, we have estimated $n_h$ for each $x$, and found a monotonic increase of $n_h$ as a function of $T_C$, as shown in Fig. 3e. We will come back to this point later.

Having established the bulk-band structure and FSs, we proceed to the evolution of Dirac-cone SS which is crucial for clarifying a possible topological phase transition. Figure 4a displays the near-$E_F$ ARPES intensity around the $\bar{\Gamma}$ point for pristine $Sb_2Te_3$ ($x = 0.0$) measured at $T = 20$ K with Xe-I photons ($hv = 8.437$ eV). Because the bulk VB is located well below $E_F$ at this $hv$ due to the $k_z$ dispersion and the intensity of Dirac-cone SS is enhanced compared to the case of $hv \sim 78$ eV shown in Figs. 2a and 3a, one can clearly see the $\Lambda$-shaped energy dispersion of the lower Dirac-cone band. It is more clearly visualized in the second-derivative plot of momentum distribution curves (MDCs) in Fig. 4b. We have investigated the electronic states slightly above $T_C$, namely in the paramagnetic (PM) phase, by tuning measurement temperatures as indicated in the $x$ vs $T$ phase diagram in Fig. 4c. As shown in Fig. 4b, d-f, the Dirac-cone SS appears to commonly exist for all the samples whereas it gradually moves upward toward unoccupied region with Cr doping. We determined the experimental band dispersion by numerically fitting the MDCs (dots in Fig. 4g) and extrapolated it by a linear function. Then, the Dirac-point energy was estimated to be ~0.03, ~0.05, ~0.08, and ~0.24 eV above $E_F$ for $T_C$=0 K, 30 K, 98 K, and 192 K samples, respectively. This signifies a systematic upward shift of the Dirac point with increasing $x$.

The fact that the Dirac-cone SS is always seen irrespective of $x$ suggests that the system maintains topologically non-trivial character at least up to the highest $x$ (0.35). Thus, one may be able to draw a plausible band diagram as a function of $x$, as illustrated in Fig. 4h. Since it is likely that the Cr doping reduces SOC and makes the system closer to the topological-phase-transition point, it is suggested that the whole bulk-band gap is gradually reduced with Cr doping while keeping the inverted bulk band structure (we will come back this point later) accompanied with a gradual filling of the spin-orbit gap at the $\bar{\Gamma}$ point. This is qualitatively reproduced by our first-principles band-structure calculations; for details see Figs. S2 and S3 of Supplementary Information (note that the reduction of bulk band gap needs to be verified by directly accessing the bulk conduction band, CB, e.g. using pump-probe ARPES). Taking into account the observed large upward shift of the Dirac-cone SS with Cr doping in Fig. 4, it is inferred that the Dirac-cone SS is gradually isolated from the VB with Cr doping, as depicted in Fig. 4h.

Now that the presence of Dirac-cone SS in wide *x* range is established, the next question is whether or not the Dirac cone affected by the ferromagnetic order can be seen by ARPES; this issue has been intensively debated in many ferromagnetic TIs and intrinsic magnetic TIs but is still highly controversial[19,20,47,48,50–57]. To answer this question, we have chosen the $T_C$=98 K sample and performed a temperature-dependent ARPES measurement. This sample is more suited to trace an influence of FM to the Dirac-cone SS compared to the $T_C$=192 K one, because larger area of the lower Dirac-cone band can be traced by ARPES while keeping reasonably high $T_C$. One can see in the second-derivative intensity plot at $T = 140$ K (above $T_C$) in Fig. 5a that $\mathbf{k}_F$ points for the lower Dirac cone at positive and negative $k_y$'s are well separated from each other (white arrows). On the other hand, such a separation is not easy to see at $T = 20$ K and the band dispersion seems *rounded* near $E_F$ (Fig. 5b). This difference in the Dirac-band dispersion is also visualized by the EDCs symmetrized with respect to $E_F$ in Fig. 5c, d that cancel out the effect of FD function (note that the symmetrization method is known to be useful even when particle-hole symmetry is broken, as long as one intends to discuss $E_F$-crossing of bands without much influence from the FD function). At $T = 140$ K (Fig. 5c), a peak feature originating from the Dirac-cone band rapidly disperses toward $E_F$ on approaching the $\bar{\Gamma}$ point, and appears to reach $E_F$ around the $\bar{\Gamma}$ point (see dashed curve). On the other hand, at $T = 20$ K, the peak feature does not reach $E_F$ and stays at ~0.05 eV at the $\bar{\Gamma}$ point, supporting the rounding behavior of the lower Dirac-cone branch. Such a critical difference is also highlighted in the symmetrized EDC at the $\bar{\Gamma}$ point in Fig. 5e in which spectral weight at $E_F$ is obviously suppressed at $T = 20$ K with respect to that at $T = 140$ K. We also found that MDCs at several binding energies shown in Fig.5f show a marked difference between $T = 140$ K and 20 K, namely, the separation between two peaks becomes smaller at $T = 20$ K. These intriguing spectral changes with temperature are likely associated with the ferromagnetic transition and not with a simple thermal-broadening effect or some other extrinsic effects because the spectral feature shows a marked change only below $T_C$ (60 and 20 K) as highlighted by the MDCs in Fig. 5g. The ferromagnetic origin of the Dirac-cone modulation is also supported by the very small temperature variation of the MDCs in the pristine $T_C$=0 K sample as shown in Fig. 5h. As schematically shown in Fig. 5i, these results can be naively explained in terms of the formation of a magnetic gap (Zeeman gap) at the Dirac point due to the time-reversal-symmetry breaking associated with the ferromagnetic order[24] but the observed change in the band structure even away from the Dirac point may not be fully explained in terms of this scenario. It is noted here that the magnetic gap in the unoccupied state is suggested from the XAS-XMCD measurements[62], in support of the present ARPES result.

What are the implications of the observed bulk-band evolution in relation to the mechanism of high-temperature FM? As stated above, plausible origins of FM are the RKKY interaction[22,32,35,36,41,42,44–46] and the van-Vleck mechanism[2,27,37,40]. In the former case, $T_C$ is proportional to the density of states at $E_F$, because the interactions among magnetic ions are mediated by itinerant conduction electrons. The gradual filling of the VB valley and expansion of elongated FS petals with Cr doping shown in Fig. 3a, b, and the experimental fact that carrier concentrations estimated from the FS volume of bulk 3D hole pockets monotonically increase on increasing $T_C$ (Fig. 3e), are consistent with the RKKY scenario. Also, $T_C$ values estimated from the RKKY model[63] based on the experimental band parameters reaches 70-80 % of experimental $T_C$ values, in favor of the RKKY scenario (for details, see Section 3 of Supplementary Information). Moreover, the fact that $T_C$ tends to become higher in the metallic samples compared to the insulating one [Bi-doped $(Cr_xSb_{1-x})_2Te_3$][24,25] supports the conclusion that carrier-mediated interaction dominates the FM. This also suggests that the carrier-independent superexchange interaction which is expected to become FM type in Cr-doped $Sb_2Te_3$ (refs. 32, 33) is unlikely responsible for the high $T_C$. In the van-Vleck mechanism, FM is stabilized by a strong enhancement of magnetic susceptibility along $c$ axis due to the band inversion[2] associated with hypothetical excitations from the valence to the CB. We suggest that the observed band structure is not in contradiction with the van-Vleck scenario because the filling of VB valley with Cr doping decreases the excitation energy around the $\bar{\Gamma}$ point (see Fig. 4h) and may lead to the enhancement of magnetic susceptibility (note that validation of this scenario requires an estimation of the energy position of the CB edge in the unoccupied state). Thus, the present results suggest a key role of RKKY interaction, possibly supported by the van-Vleck mechanism, to stabilize FM at such high $T_C$.

The observed persistence of topological phase in wide range of $x$ has important implications in the band engineering of magnetic TIs. We found in Fig. 3 that the bulk bands remain inverted even when considerable amount (maximally 35%) of Sb atoms are replaced with Cr atoms. Such robustness against chemical substitution with weaker SOC element may be largely associated with the strong SOC of the Te 5$p$ orbital. This speculation is supported by (i) the first-principles band-structure calculations of tetradymites where the band inversion of the Te compound is more robust against chemical substitution than that of the Se counterpart[27,28], and by (ii) the experimental fact that the replacement of Se with Te easily leads to topological phase transition at the fixed Cr content $x$[27]. An important aspect of our finding is that the band structure near $E_F$ is dramatically modified while always keeping the topological character. This points to an intriguing possibility that the band-structure modulation (i.e. flattening of the VB top) itself can assist/trigger the robust FM in TIs. This is distinct from the previous band-engineering studies on the chemically substituted TIs that are mainly aimed at the control of band-gap magnitude[60,64] and the location of chemical potential[65,66].

Finally, we discuss the implications of the unusual temperature evolution of Dirac-cone SS. The apparent change in the Dirac-cone dispersion across $T_C$ revealed in Fig. 5 strongly suggests the robust (and likely uniform) nature of FM in (Cr$_x$Sb$_{1-x}$)$_2$Te$_3$ and the presence of strong exchange interaction between Cr 3$d$ and Dirac electrons. The change in the Dirac-cone dispersion across $T_C$ in surprisingly wide energy range (at least up to 0.2 eV with respect to the Dirac point) is distinct from the general expectation that the FM just modifies the Dirac-band structure within small energy scale around the Dirac point[19,20,50,54–57]. This suggests that ferromagnetic order influences the whole Dirac-cone band in (Cr$_x$Sb$_{1-x}$)$_2$Te$_3$. To clarify the microscopic origin of this finding, a detailed investigation of spin texture with higher resolution would be necessary. It is emphasized here that a coupling between spontaneous FM and Dirac electrons inferred from the present study is a direct benefit of the coexistence of high-temperature FM and topological nature. Such a strong coupling in spin-helical Dirac electrons is potentially useful to manipulate magnetization of magnetic TIs *via* current-induced spin accumulation and resultant high spin-orbit torque[7-12]. The large energy scale of magnetic interaction would be also useful to develop the low-power-consumption topological spintronics devices working at higher temperatures.

**METHODS**

Sample preparation

High-quality thin films of $(Cr_xSb_{1-x})_2Te_3$ with typical thickness of 150 nm were grown on semi-insulating GaAs(111)-B substrate using the molecular-beam-epitaxy (MBE) technique. The Cr composition $x$ and $T_C$ were determined by energy-dispersive X-ray spectroscopy (EDX) and a vibrating sample magnetometer, respectively. Details of the sample fabrication and characterization were described elsewhere[26].

ARPES and XMCD measurements

ARPES measurements were carried out using a Scienta-Omicron DA30 spectrometer with the beamline BL-28A at Photon Factory, KEK. Circular polarized light with photon energy ($hv$) of 60–100 eV was used to excite photoelectrons. ARPES measurements were also carried out using an MBS-A1 electron-energy analyzer with a high-flux xenon discharge lamp ($hv$ = 8.437 eV) at Tohoku University. XMCD measurements were carried out at Dragon Beamline (BL 11A) of the Taiwan Light Source, and total electron yield method was used[67]. For the ARPES and XMCD measurements, we evaporated Te capping layer of 50 nm on top of the $(Cr_xSb_{1-x})_2Te_3$ film after the growth and then extracted the sample from the MBE chamber. Next we mounted an aluminum post on top of the Te capping layer, installed the sample into the ARPES/XMCD chamber, and then cleaved it under ultrahigh vacuum to obtain a clean surface.

First-principles band-structure calculations

First-principles band-structure calculations were carried out by using the Quantum Espresso software package[67] within local density approximations (LDA)[68] and LDA+$U$ calculations[69] in which on-site Coulomb interaction for Cr 3$d$ orbital state was considered with the $U$ value of 3.0 eV. SOC was taken into account with fully relativistic ultrasoft pseudopotentials. The cutoff energy was set to be 60 Ry and 500 Ry for plane-wave basis set and charge density, respectively. The electronic properties were calculated by using 12×12×1 $k$-points mesh after the atomic structures were optimized. To calculate the band structure of Cr-17%-replaced $(Cr_{1/6}Sb_{5/6})_2Te_3$, we have assumed the slab structure where two Sb sites among six Sb sites in 15-atom unit cell are replaced with Cr atoms.

**ACKNOWLEDGMENTS**

We thank M. Kitamura, K. Horiba, and H. Kumigashira for their assistance in the ARPES experiments. We also thank T. Dietl for fruitful discussions. This work was supported by JST-CREST (No: JPMJCR18T1), JST-PRESTO (No: JPMJPR18L7), JSPS (JSPS KAKENHI No: JP17H01139, JP19H01845), the Center for Spintronics Research Network at Tohoku University, KEK-PF (Proposal number: 2018S2-001, 2021S2-001), and UVSOR (Proposal number: 30-858, 30-846), Ministry of Science and Technology of the Republic of China, Taiwan under contract no. MOST 108-2112-M-213-001-MY3. C.W.C and Y.N. acknowledge GP-spin and JSPS for financial support.

**AUTHOR CONTRIBUTIONS**

The work was planned and proceeded by discussion among K.H. S.S., F.M. A.C. and T.S. S.G. and F.M. fabricated thin films and characterized them. C.W.C., Y.N., F.H.C., H.J.L., C.T.C., and A.C. performed the XMCD measurements. T.P.T.N. and K.Y. carried out the band-structure calculations. K.H., S.S., K.N., T.T., and T.S. performed the ARPES measurements. K.H., S.S., and T.S. finalized the manuscript with inputs from all the authors.

**ADDITIONAL INFORMATION**

**Correspondence and requests** for materials should be addressed to S. S. and T. S. (e-mail: s.souma@arpes.phys.tohoku.ac.jp and t-sato@arpes.phys.tohoku.ac.jp)

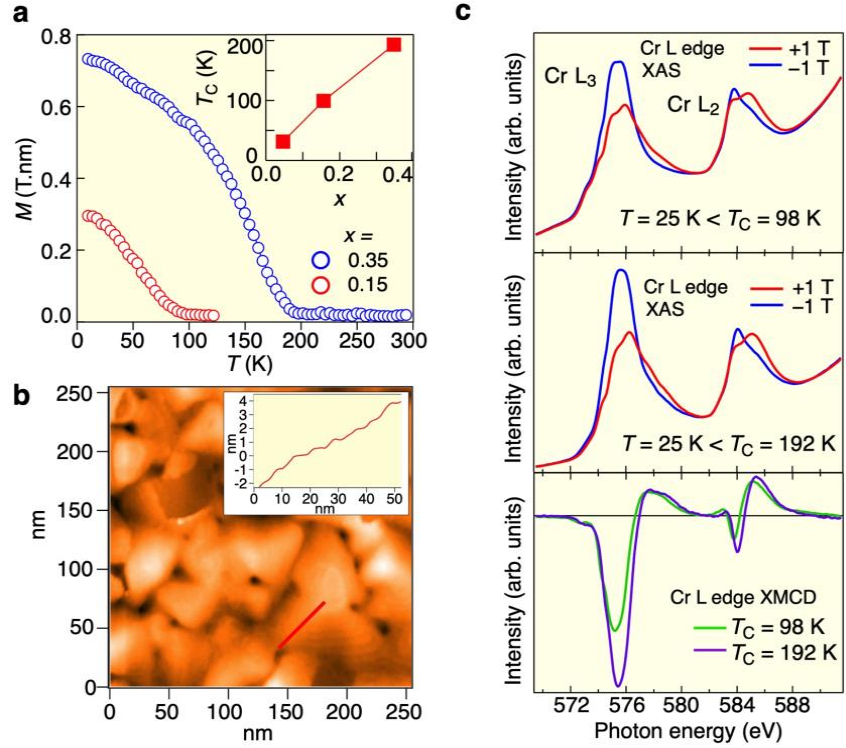

**FIG. 1. Characterization of $(Cr_xSb_{1-x})_2Te_3$ films.** **a**, Temperature-dependent magnetization of $(Cr_xSb_{1-x})_2Te_3$ films at Cr concentrations of $x$ = 0.15 and 0.35. $T_C$ was estimated as 98 and 192 K, respectively. Magnetic field of 2 mT was applied along the $c$ axis (easy axis) of the crystal. Inset shows $x$-dependence of $T_C$ estimated from the magnetization measurements[24]. **b**, Atomic-force-microscopy image on the cleaved surface of $Sb_2Te_3$. Inset shows a line profile along red line in **b**. **c**, Magnetic-field dependence of X-ray absorption spectra around the Cr $L_2$ and $L_3$ edges for the (top) $T_C$ =98 K sample and (middle) $T_C$=192 K sample, measured at $T$ = 25 K. (Bottom) comparison of XMCD intensity at 25 K between the two samples. Magnetic field was applied perpendicular to the film (along the $c$-axis).

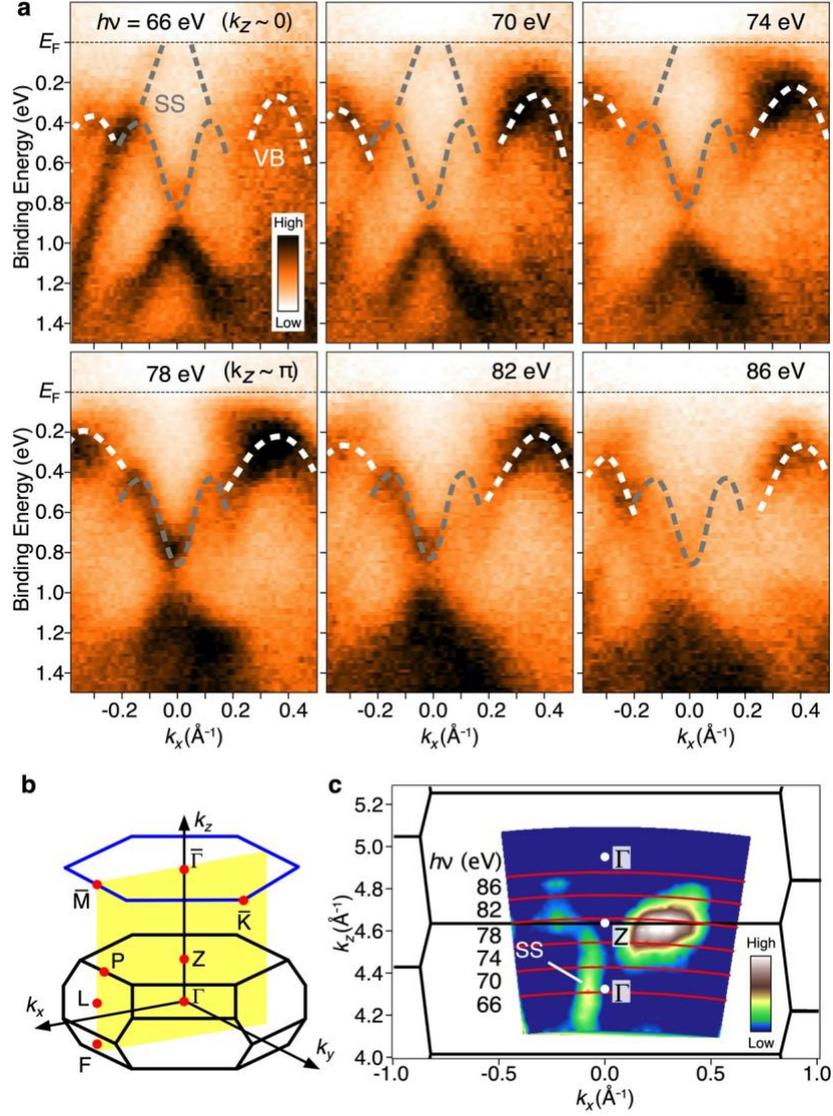

**FIG. 2. 3D band structure of Sb₂Te₃. a**, Photon-energy dependence of the ARPES intensity for pristine Sb$_2$Te$_3$ measured along the $\overline{\Gamma}\overline{M}$ cut of surface BZ which signifies a clear $k_z$ dispersion of the bulk VB. **b**, Bulk rhombohedral BZ (black) and surface hexagonal BZ (blue) of (Cr$_x$Sb$_{1-x}$)$_2$Te$_3$. **c**, ARPES-intensity mapping at $E_F$ in the $k_x$-$k_z$ plane for pristine Sb$_2$Te$_3$ measured at $T = 30$ K by varying $h\nu$. ARPES intensity at $E_F$ was obtained by integrating the spectral intensity within ± 30 meV of $E_F$.

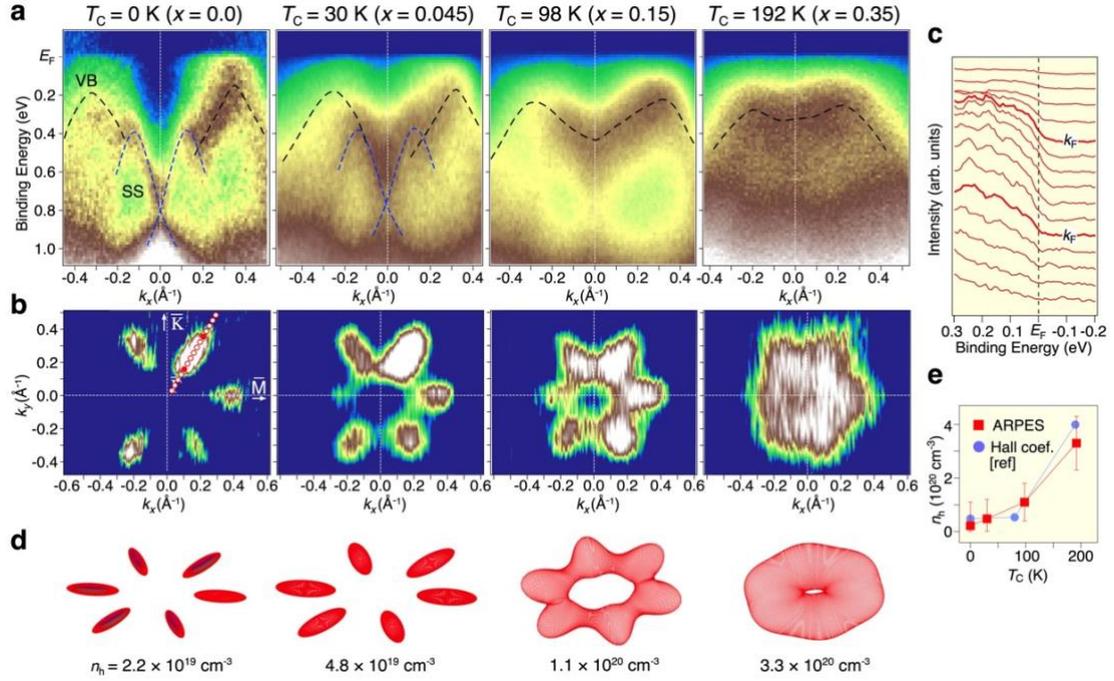

**FIG. 3. Evolution of band structure upon Cr doping. a**, **b** ARPES intensity along the $k_x$ cut ($\overline{\Gamma}\overline{M}$ cut) and FS mapping, respectively, measured at $T = 30$ K with $h\nu = 78$ eV for $(Cr_xSb_{1-x})_2Te_3$ ($x = 0.0$, 0.045, 0.15, and 0.35). Blue dashed and black dashed curves in **a** are guides for the eyes to trance the band dispersion of SS and VB, respectively. **c**, EDCs along a **k** cut crossing the elongated hole pocket for $x = 0$ (red open circles in **b**; red filled circles correspond to the $k_F$ points). We defined location of the $k_F$ point as the **k** point at which the Fermi-edge intensity is reduced by 20% with respect to the maximum value. **d**, Schematic 3D FS as a function of Cr concentration estimated by fitting $k_F$ points of each sample in **b**. **e**, Carrier concentrations (red squares) plotted against $T_C$ estimated from the volume of experimental bulk FS for each $x$. Carrier concentrations estimated from the Hall-conductivity measurements[24] are also overlaid by blue circles.

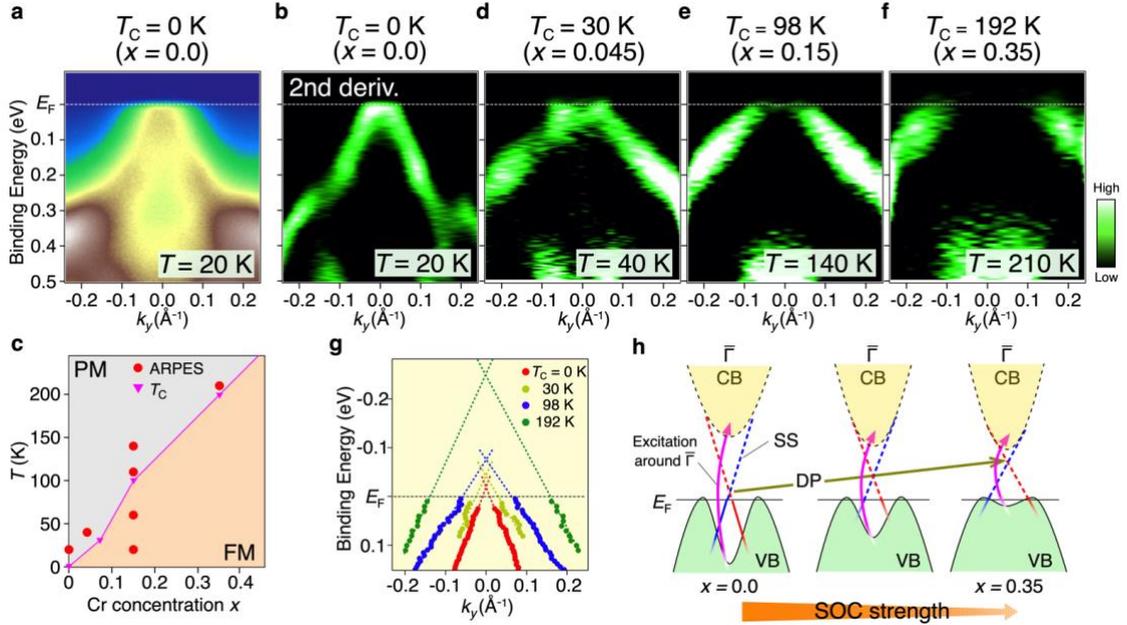

**FIG. 4. Evolution of Dirac-cone SS upon Cr doping. a, b** Near-$E_F$ ARPES intensity and corresponding second-derivative intensity plot of the MDCs, respectively, for $Sb_2Te_3$ ($T_C = 0$ K) measured along the $k_y$ cut at $T = 20$ K with the Xe-I line ($h\nu = 8.437$ eV). **c**, Magnetic phase diagram as a function of Cr concentration $x$, together with the temperature at which the measurements shown in Figs. 4 and 5 were performed. **d-f**, Second-derivative intensity plot of the MDCs along the $k_y$ cut measured at slightly above $T_C$ of each sample with the Xe-I line for $x = 0.045$, 0.15, and 0.35. **g**, Experimental Dirac-cone dispersion (dots) estimated from the numerical fittings to the MDCs for $x = 0.0$, 0.045, 0.15, and 0.35. Dashed lines are a linear extrapolation of the Dirac-cone bands. **h**, Schematic band diagram of the evolution of bulk and surface bands in $(Cr_xSb_{1-x})_2Te_3$ obtained from the present ARPES result. The CB likely moves downward upon increasing $x$ because the system approaches to the topological phase transition point.

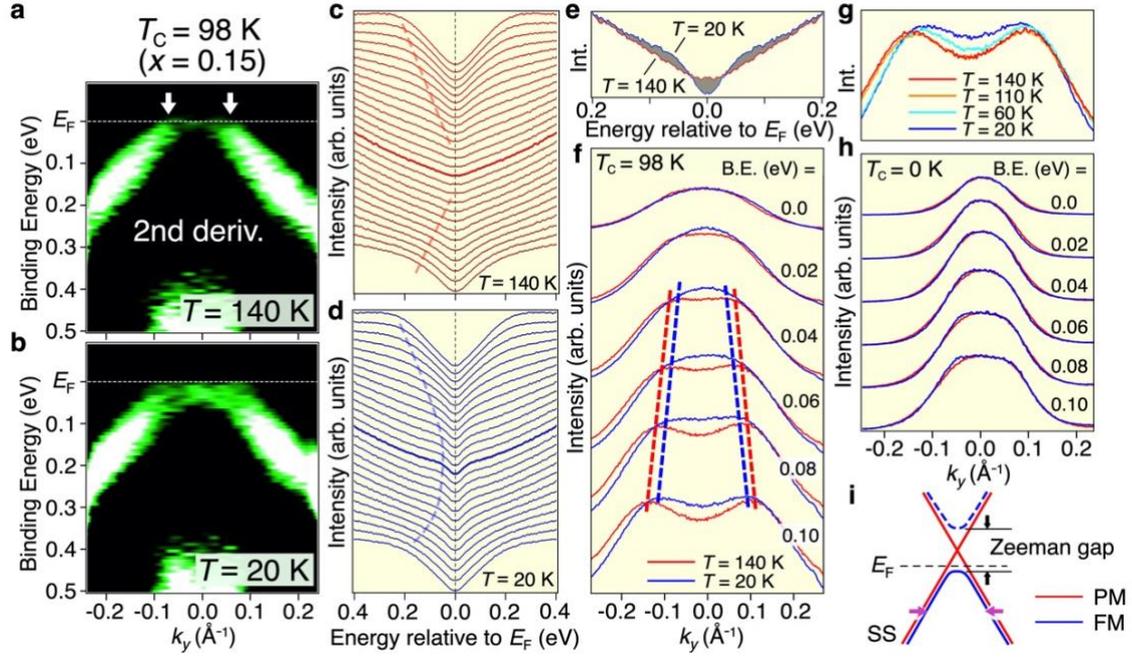

**FIG. 5. Change in the Dirac-cone dispersion across $T_C$. a, b** Second-derivative intensity plot of the MDCs for $(Cr_xSb_{1-x})_2Te_3$ ($x = 0.15$) along the $k_y$ cut measured at $T = 140$ K (above $T_C$) and 20 K (below $T_C$), respectively. White arrows in **a** indicate the $\mathbf{k}_F$ points. **c, d** EDCs symmetrized with respect to $E_F$ along the $k_y$ cut at $T = 140$ K and 20 K, respectively. **e** Direct comparison of the EDCs at the $\bar{\Gamma}$ point at $T = 140$ K and 20 K. Shaded area highlights the spectral-weight difference. **f,** Direct comparison of the MDCs at representative $E_B$ slices measured at $T = 140$ and 20 K. **g,** Temperature dependence of the MDCs at $E_B = 0.1$ eV. **h,** Same as **f** but for pristine $Sb_2Te_3$. **i,** Schematic illustration of the band dispersion of the Dirac-cone SS above and below $T_C$ for $x = 0.15$, indicated by red and blue curves, respectively.

# Supplementary information for "Unusual band evolution and persistence of topological surface states in high-$T_C$ magnetic topological insulator"

K. Hori[1], S. Souma[2,3], C.-W. Chuang[1], Y. Nakata[1], K. Nakayama[1], S. Gupta[4,†], T. P. T. Nguyen[5], K. Yamauchi[5], T. Takahashi[1], F. Matsukura[4,6], F. H. Chang[7], H. J. Lin[7], C. T. Chen[7], A. Chainani[7], and T. Sato[1,2,3,8,9]

[1]*Department of Physics, Graduate School of Science, Tohoku University, Sendai 980-8578, Japan*

[2]*Center for Science and Innovative in Spintronics, Tohoku University, Sendai 980-8577, Japan*

[3]*Advanced Institute for Materials Research (WPI-AIMR), Tohoku University, Sendai 980-8577, Japan*

[4]*Center for Innovative Integrated Electronic Systems, Tohoku University, Sendai 980-0845, Japan*

[5]*Department of Precision Engineering, Graduate School of Engineering, Osaka University, 2-1 Yamadaoka, Suita, Osaka 565-0871, Japan*

[6]*RIKEN Center for Emergent Matter Science, Wako, 351-0198*

[7]*National Synchrotron Radiation Research Center, Hshinchu, 30077, Taiwan ROC*

[8]*International Center for Synchrotron Radiation Innovation Smart (SRIS), Tohoku University, Sendai 980-8577, Japan*

[9]*Mathematical Science Center for Co-creative Society (MathCCS), Tohoku University, Sendai 980-8577, Japan*

[†]*Present address: Department of Physics, Bennett University, Greater Noida, 201310, India*

**Section 1: Cr-doping evolution of band structure for $(Cr_xSb_{1-x})_2Te_3$**

To demonstrate unusual evolution of bulk-band structure as a function Cr doping, we show in Fig. S1a ARPES intensity along the $k_x$ cut ($\bar{\Gamma}\bar{M}$ cut) measured at $T = 30$ K with $hv = 78$ eV for $(Cr_xSb_{1-x})_2Te_3$ ($x = 0.0$, $0.045$, $0.15$, and $0.35$; same as Fig 3a in the main text), together with the experimental energy dispersion of the bulk band (red circles) estimated from the peak position of EDCs/MDCs. One can see deep valley at the $\bar{\Gamma}$ point for $x = 0.0$ whereas the valley is gradually filled in upon increasing $x$. This unusual filling-in behavior can be directly identified by comparing EDCs at the $\bar{\Gamma}$ point (Fig. S1b) and away from the $\bar{\Gamma}$ point ($k_x = 0.2$ Å$^{-1}$; Fig. S1c). At the $\bar{\Gamma}$ point, the peak position $E_P$ (vertical bar) shifts by ~ 0.4 eV between $x = 0.0$ and 0.35 ($E_P$ ~ 0.7 eV for $x = 0.0$ and ~ 0.3 eV for $x = 0.35$) whereas the total shift is reduced to ~0.2 eV at away from the $\bar{\Gamma}$ point ($E_P$ ~ 0.4 eV for $x = 0.0$ and ~ 0.2 eV for $x = 0.35$), signifying a non-rigid-band-like band evolution upon Cr doping.

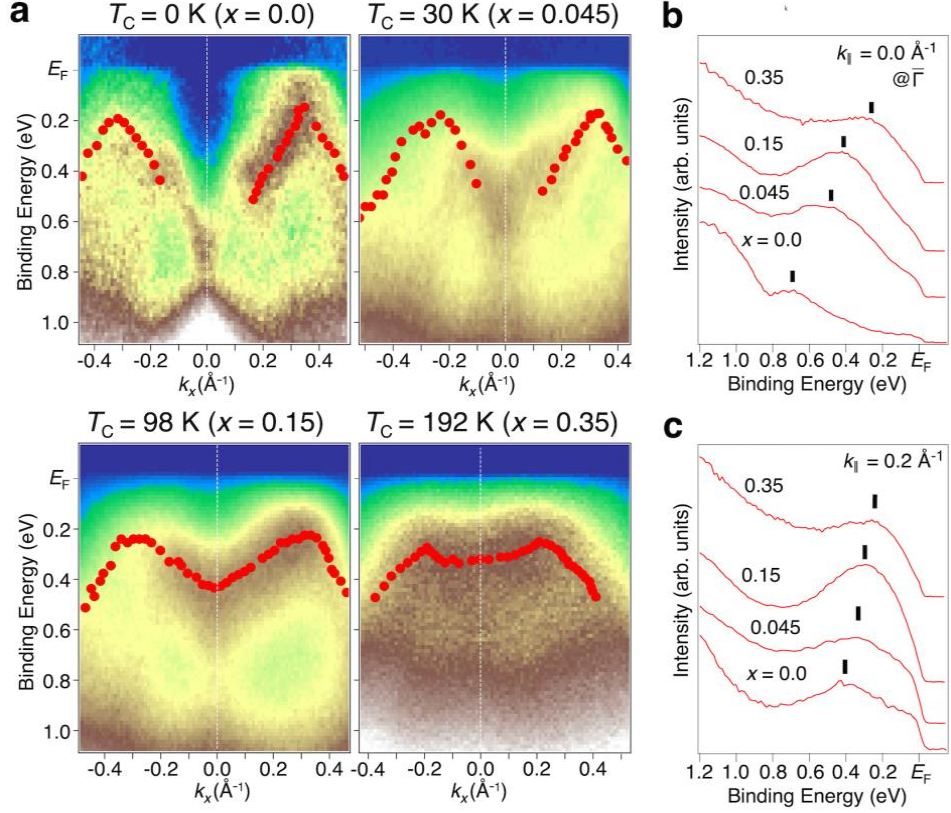

**Fig. S1: Unusual filling-in behavior of topmost valence band upon Cr doping.** **a**, ARPES intensity along the $k_x$ cut ($\overline{\Gamma}\overline{M}$ cut) measured at $T = 30$ K with $h\nu = 78$ eV for $(Cr_xSb_{1-x})_2Te_3$ ($x = 0.0$, 0.045, 0.15, and 0.35), together with the experimental bulk band dispersion estimated from the peak position of EDCs/MDCs. **b**, **c** $x$ dependence of EDC at and away from the $\overline{\Gamma}$ point ($k_x = 0.0$ and $0.2$ Å$^{-1}$), respectively. Vertical bar corresponds to the peak position.

### Section 2: Calculated band structure for pristine and Cr-doped Sb₂Te₃

To support the evolution of experimental band structure upon Cr doping, we have carried out bulk-band structure calculations for pristine $Sb_2Te_3$ and Cr-17%-replaced $(Cr_{1/6}Sb_{5/6})_2Te_3$ with the slab structure in which one Sb sites among six Sb sites in 15-atom unit cell are replaced with Cr atoms[S1], as seen in Fig. S2a, b. One can recognize from a comparison of band structure between pristine $Sb_2Te_3$ (Fig. S2c) and $(Cr_{1/6}Sb_{5/6})_2Te_3$ (Fig. S2d) that the band gap around the $\Gamma$ point is reduced upon Cr substitution, in line with the conclusion drawn from the ARPES result. Moreover, the energy level around the A point appears to move up with respect to the top of the VB at the midway between the A and L points, qualitatively consistent with the experimental observation of the filling-in behavior seen in Fig. 3 of the main text. To better visualize this trend, we show in Fig. S3 direct comparison of experimental band structure for $x =$ 0.0 and 0.15 compared with the corresponding calculated band structure for $Sb_2Te_3$ and $(Cr_{1/6}Sb_{5/6})_2Te_3$. One can see that both the experimental and calculated topmost VB commonly show filling-in behavior upon Cr doping, suggesting that the observed evolution of band structure upon Cr doping is an intrinsic nature of $(Cr_xSb_{1-x})_2Te_3$.

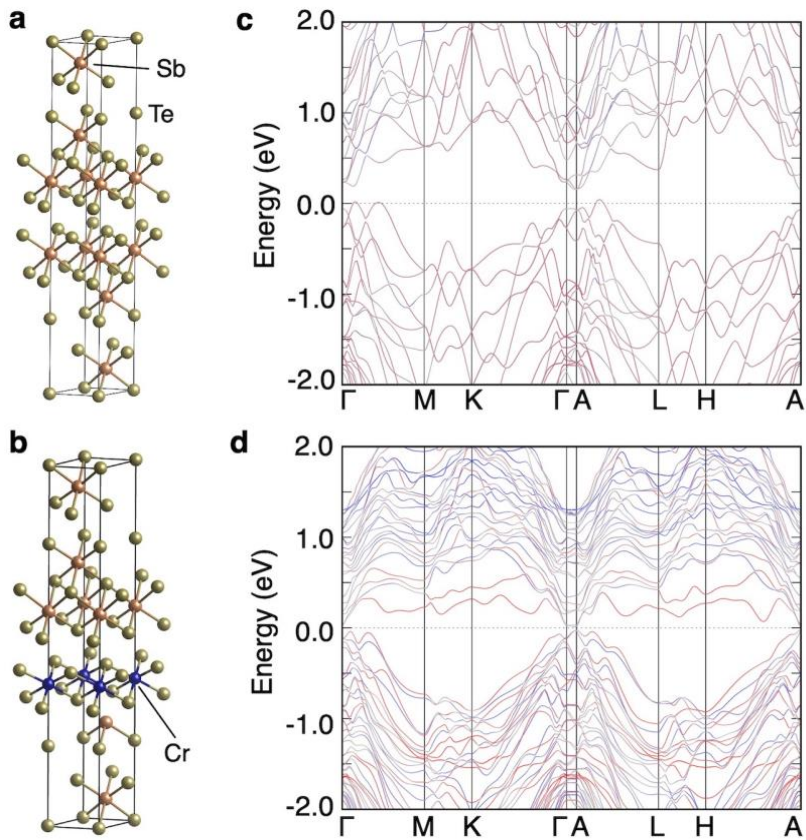

**FIG. S2: Calculated band structures for pristine and Cr-doped $Sb_2Te_3$. a, b** Crystal structure of pristine $Sb_2Te_3$ and $(Cr_{1/6}Sb_{5/6})_2Te_3$ with the hexagonal setting, adopted for calculating the band structure. We used the experimental lattice constants of $Sb_2Te_3$ ($a =$ 4.25 Å; $c =$ 30.35 Å) for both structures. Red and blue curves represent spin-up and spin-down bands, respectively.

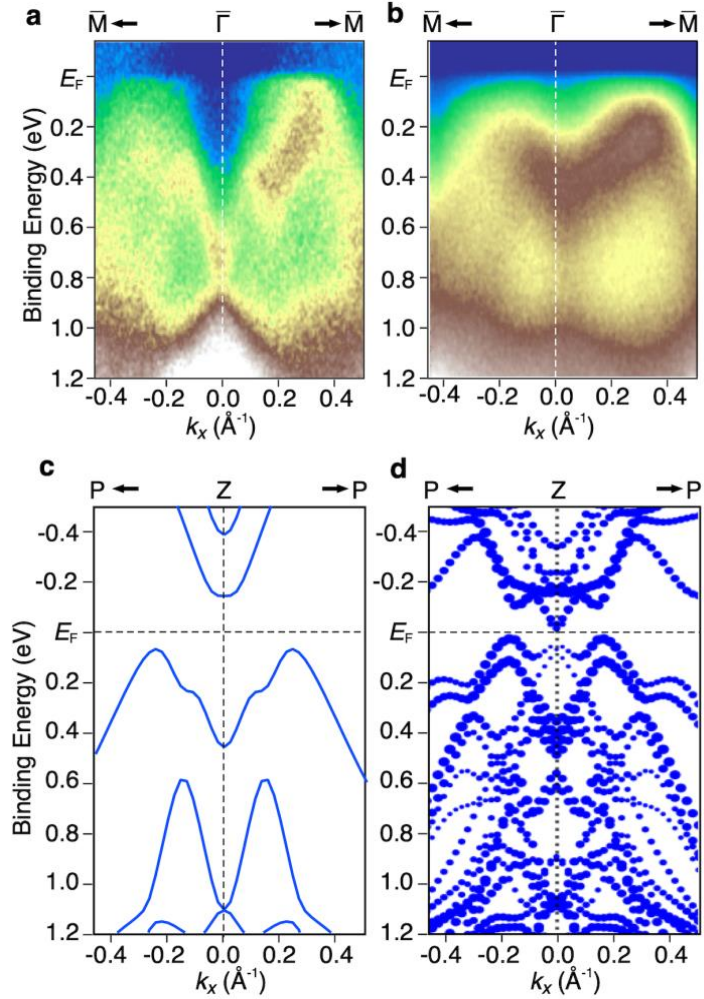

**FIG. S3: Comparison of experimental and calculated band structure. a, b** ARPES intensity along the $k_x$ cut ($\overline{\Gamma M}$ cut) measured at $T = 30$ K with $h\nu = 78$ eV for $x = 0.0$ and 0.15, respectively (same as Fig. 3a in the main text). **c, d**, Calculated band structure along the same cut at $k_z = \pi$ (ZP cut in the rhombohedral BZ) for pristine $Sb_2Te_3$ and $(Cr_{1/6}Sb_{5/6})_2Te_3$, respectively. The calculated bands for hexagonal BZ in Fig. S2 are unfolded into the first zone for rhombohedral BZ. Size of circles in **d** reflects the spectral weight.

**Section 3: Estimation of $T_C$ based on the RKKY model**

To estimate the contribution of hole-carrier mediated interaction to the FM of $(Cr_xSb_{1-x})_2Te_3$, we have estimated $T_C$ based on the RKKY model[S2], by using the following equation,

$$k_B T_C = \frac{x}{12}(N_0\beta)^2 S(S+1)\frac{D(E_F)}{N_a} \quad (1)$$

where $k_B$ is a Boltzmann constant, $N_0, \beta, S, D(E_F), N_a$ are density of cation, $p$-$d$ exchange constant, spin of magnetic ion, density of states at $E_F$ with the unit of states/eV/unit-cell, number of cation per unit cell, respectively. Here, $S = 3/2$ for $Cr^{3+}$ ion. We have estimated $N_0\beta$ to be 0.74 eV by referring to the structural parameter of $Sb_2Te_3$ (ref. S2). To estimate $D(E_F)$, we used the following equation,

$$D(E_F) = \frac{2V_{uc}}{(2\pi)^3}\iint_{S:E(\mathbf{k})=0} dS \frac{1}{|\nabla_\mathbf{k} E(\mathbf{k})|} \quad (2)$$

where $V_{uc}$ is the unit cell volume, $dS$ is the element of surface area in $k$ space. Since the FS for $x = 0.15$ and $0.35$ has a modulated donut shape as schematically shown in Fig. S4a, we have approximated the dispersion relationship by assuming the following equation in cylindrical coordinate system

$$E = \frac{\hbar^2}{2m_b}\left(k_r - k_c(\theta)\right)^2 + \frac{\hbar^2}{2m_c}k_z^2 + E_0(\theta) \quad (3)$$

where the vertical slice in the constant $\theta$ plane at fixed energy $E$ show ellipsoidal shape as shown by blue ellipsoids in Fig. S4a. Here, $m_b$ and $m_c$ represent effective mass along the longer and shorter axes of the ellipsoid, $k_r$, $k_c$, and $E_0(\theta)$, $k$ distance from the $k_z$ axis, center of ellipsoid, and top of the hole band at given $\theta$, respectively. We have assumed the radius of inner hole of donut to be constant $(k_i = a)$ and outer radius of donut to be $k_o = b + c\cos 3\theta + d\cos 6\theta$ to satisfy the crystal symmetry, where the center of ellipsoid is $k_c = (k_o + k_i)/2$ (see Fig. S4b). By numerically fitting experimental $\mathbf{k}_F$ points at $k_z = \pi$ plane, we have estimated the $a, b, c, d,$ and $m_b$ values. For the band mass along $k_z (m_c)$, we referred to that of topmost VB in our DFT calculations for $Sb_2Te_3$. Then, we have estimated hole carrier concentration $n_h$ from the FS volume and $D(E_F)$ from the gradient of energy dispersion. Consequently, we estimated $T_C$ to be 66 K and 152 K for $x = 0.15$ and $0.35$, respectively (Table S1), which is about 70-80 % of the experimental values. This suggests a dominant role of RKKY interaction to enhance $T_C$ in $(Cr_xSb_{1-x})_2Te_3$ of highly-doped region.

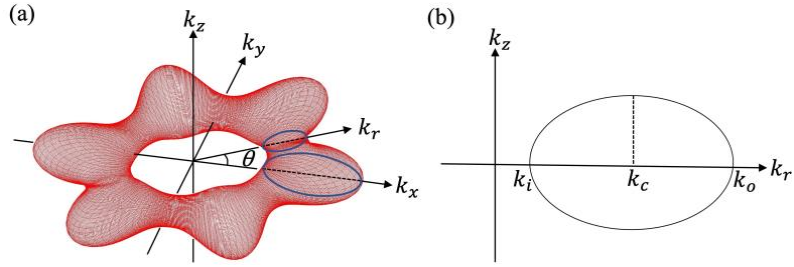

**Fig. S4: 3D FS for higher $T_C$ samples. a,** Assumed modulated donut FS for $x = 0.15$. Definition of $\theta$ is also shown. **b**, Slice of FS at constant $\theta$ ($k_r$ - $k_z$ plane) together with the definition of $k_i$, $k_c$, and $k_o$.

|  | $x = 0.15$ | $x = 0.35$ |
|---|---|---|
| $T_C$ (expt.) [K] | 98 | 192 |
| $T_C$ (calc.) [K] | 66 | 152 |
| $D(E_F)$ [states/eV/u.c.] | 1.3 | 1.3 |
| $n_h$ [$10^{20}$ cm$^{-3}$] | 1.1 | 3.3 |
| $m_b$ | $-0.5\ m_0$ | $-0.65\ m_0$ |
| $m_c$ | $-0.2\ m_0$ | $-0.2\ m_0$ |

**Table S1: Estimation of $T_C$ and band parameters.** Table of experimental $T_C$, estimated $T_C$, $D(E_F)$, $n_h$, $m_b$, and $m_c$ for $x = 0.15$ and 0.35. We take hexagonal unit cell for (Cr$_x$Sb$_{1-x}$)$_2$Te$_3$ as shown in Fig. S2a.

**Supplementary references**

[S1] Ruan, Y. *et al.* Band-structure engineering of the magnetically Cr-doped topological insulator Sb$_2$Te$_3$ under mechanical strain. *J. Phys. Condens. Matter* **31**, 385501 (2019).

[S2] Dietl, T., Ohno, H. & Matsukura, F. Hole-mediated ferromagnetism in tetrahedrally coordinated semiconductors. *Phys. Rev. B* **63**, 195205 (2001).